\documentclass[twocolumn,showpacs]{revtex4-1}
\usepackage[T2A]{fontenc}
\usepackage[cp1251]{inputenc}
\usepackage[english]{babel}
\usepackage{amsmath}
\usepackage{amsfonts}
\usepackage{amssymb}
\usepackage{graphicx}
\usepackage{color}
\usepackage{bm}

\renewcommand{\cosh}{\mathop{\rm ch}}

\newcommand{\e}{\mathrm{e}}

\renewcommand{\i}{\mathop{\rm i}}
\renewcommand{\d}{\mathrm d}
\renewcommand{\url}{}

\newcount\timehh  \newcount\timemm
\timehh=\time \divide\timehh by 60
\timemm=\time
\renewcommand{\i}{\mathrm i}
\renewcommand{\d}{\mathrm d}

\begin{document}

\title{Spin coherence generation and detection in spherical nanocrystals}
\author{M.M. Glazov}
\affiliation{Ioffe Physical-Technical Institute of the RAS, 194021
  St-Petersburg, Russia} 

\author{D.S. Smirnov}
\affiliation{Ioffe Physical-Technical Institute of the RAS, 194021
  St-Petersburg, Russia} 


\begin{abstract}
Theoretical description of electron spin orientation and detection by
short optical pulses is proposed for the ensembles of the singly
charged semiconductor nanocrystals. The complex structure of the
valence band in spherical nanocrystals is taken into account. We
demonstrate that the direction of electron spin injected by the pump pulse
depends on both the pump pulse helicity and the
pump pulse power. It is shown that the train of the optical pulses can
lead to the complete orientation of the resident electron spin. The
microscopic theory of the spin Faraday, Kerr and ellipticity effects
is developed and the spectral sensitivity of these signals is
discussed. We show that under periodic pumping the pronounced
mode-locking of electron spins takes place and manifests itself as
significant spin signals at negative delays between pump and probe pulses. 
\end{abstract}


\maketitle

\section{Introduction}

One of the most important tasks of semiconductor
  spintronics, a novel branch of the condensed matter physics aimed at
  the fundamental and applied research of the charge carriers spin
  dynamics, is the study of electron and hole spin control by
  nonmagnetic
  means~\cite{dyakonov_book,ssc:optor,af:nat,wu_review_2010}. In this
  regard, the manipulation of electron spins by short optical
pulses attracts a lot of research attention
nowadays~\cite{ramsay08,phelps:237402,Greilich2009,zhukov10}, see Ref. \cite{Ramsay} for
review.

The spin control is usually realized using the
  pump-probe technique where a strong circularly polarized pump
  pulse 
orients spins of electrons, holes and their complexes and a weak
linearly polarized probe pulse monitors their spin polarization via
spin (magneto-optical) 
Faraday, Kerr and ellipticity
effects as shown schematically in
Fig.~\ref{scheme}(a)~\cite{PhysRevLett.55.1128,Zheludev1994823}. Various
aspects of the pump-probe technique and features of electron spin
dynamics manifested in the pump-probe experiments are reviewed
in~\cite{dyakonov_book,ssc:optor,Korn2010415,glazov:review}.

Among rich variety of solid-state systems where the pump-probe
technique was successfully applied, the structures with singly charged
quantum dots are of special importance~\cite{dyakonov_book,glazov:review,mikkelsen07,atature07}. Ultra-long spin relaxation
times in these systems allow one to observe various interesting
phenomena including spin precession
mode-locking~\cite{A.Greilich07212006} and nuclei-induced spin
precession frequency focusing phenomena~\cite{A.Greilich09282007}. Due
to these effects about a million of electron spins localized in
different dots are coherent and precess synchronously. In most of
pump-probe spin dynamics studies the self-assembled quantum dots (quantum disks),
where due to the size quantization and strain the ground valence band state is two-fold
degenerate and corresponds to the hole spin projections $\pm 3/2$ onto
the growth axis, were used. For such structures a microscopic theory of spin
Kerr, Faraday and ellipticity effects was developed in Ref.~\cite{yugova09}. This theory is
in a good agreement with
experiments~\cite{glazov2010a,glazov:review}.

The aim of the present paper is to address theoretically the
processes of the spin coherence generation, control and detection in
spherical nanocrystals (NCs) where the valence band is four-fold
degenerate. The specifics of the optical selection rules in this case
results in novel qualitative features of spin generation and detection
processes absent in the quantum disks. In particular, as shown below,
 the direction of electron  spin
initiated by a single pump pulse or by the pump pulse train depends on
the pump pulse power: by changing the power of the circularly polarized
pump pulse one can generate electron spin oriented either along or in
the opposite direction with respect to the light propagation axis. We
demonstrate also that the excitation of electron spins by the periodic
train of pump pulses can result in the complete orientation of
electron spin and in the spin coherence mode-locking if the transverse
magnetic field is applied.

The paper is organized as follows. In Sec.~\ref{sec2} we provide a
theoretical description of spin coherence generation in a single
nanocrystal under circularly polarized pump pulse. Then in
Sec.~\ref{sec3} we analyze spin dynamics in the external magnetic
field and discuss the spin accumulation processes, among them the spin
coherence mode-locking. Section~\ref{sec4} is divided into two parts,
first one is aimed to provide a description of spin Faraday, Kerr
and ellipticity signals formation, and the second one presents temporal
dependences of spin signals in the nanocrystal arrays. The results are
summarized in the Sec.~\ref{sec5}.


\section{Spin coherence generation}
\label{sec2}
An array of spherical NCs grown of III-V compounds is considered. The
NCs are assumed to be singly charged with electrons. The ground state
of the dot corresponds to the electron at the lowest single
size-quantization level, this state is twofold degenerate with respect
to the electron spin projection on a given axis. The excited state we
consider is populated as a result of the pump pulse action; it is the
singlet trion state, which consists of a pair of electrons with
anti-parallel spins and a hole. Hereafter we assume that the optical
(carrier) frequencies of the pump, $\omega_p$, and probe,
$\omega_{pr}$, pulses are close to the singlet trion resonance
frequency, $\omega_0$, and we neglect therefore all other excited
states in the system, e.g. the triplet trion. 

\begin{figure}[t]
\includegraphics[width=\linewidth]{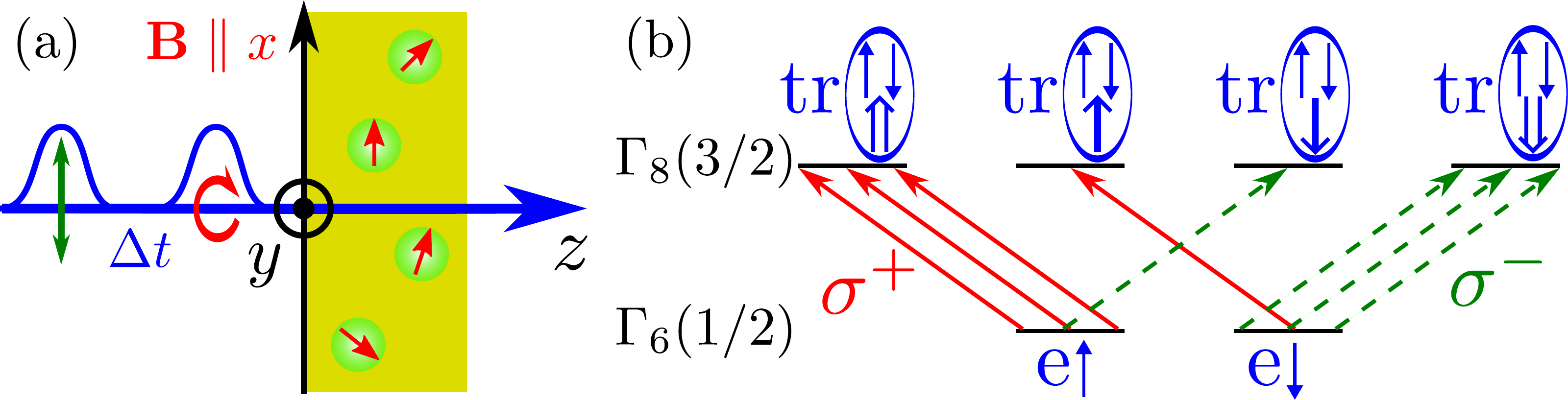}
\caption{(a) The principal scheme of the pump-probe measurements
  technique. First (pump) pulse is circularly polarized, and the
  weak probe pulse is linearly polarized, it arrives at a
    sample delayed by  $\Delta t$ with respect to the pump pulse. 
(b) Scheme of optical transitions in the case of degenerate valence
band in the charged NC. Transitions under $\sigma^+$ ($\sigma^-$)
polarized pulse are shown 
by solid (dashed) arrows. 
}\label{scheme}
\end{figure}

The singlet trion state degeneracy is determined by the hole spin
states. In spherical NCs under study the ground state of the hole
transforms according to the $\Gamma_8$ representation of $T_d$ point
symmetry group. We assume that the radius of the NC $R$ is small
enough, $R \ll a_B$, where $a_B$ is exciton Bohr radius, and neglect
the Coulomb interaction. Therefore the trion state is fourfold
degenerate, and its states can be labelled as $F_z=\pm3/2$,
$\pm1/2$. In the spherical (isotropic) approximation $F_z$ is the
component of the hole total angular momentum, which includes the
orbital momentum of size-quantized state, $L$, and the angular
momentum of Bloch function,
$J$~\cite{gelmont-dyakonov,baldereschi73,rodina89,goupalov98}. The
singlet trion wave function can be written as a product of
two-electron function 
\begin{align} 
\psi_{ee}&(\mathbf r_{e1},\mathbf r_{e2}) = \nonumber \\
&f_e(\mathbf{r}_{e1})f_e(\mathbf{r}_{e2})\frac{\left|1/2\right\rangle_1\left|-1/2\right\rangle_2-\left|-1/2\right\rangle_1\left|1/2\right\rangle_2}{\sqrt{2}}, \nonumber
\end{align}
where $f_e(\mathbf{r})=({2\pi r^2R})^{-1/2}\sin{(\pi r/R)}$
is the size quantization function of the electron in the limit of
infinite barriers, $\left|s_z\right\rangle_i$ are the corresponding spinors, and a hole wave function 
\begin{align} 
\psi_{h,F_z}(\bm r_h)=f_0(r_h)\left|L=0, J=3/2, F=3/2,
  F_z\right\rangle + \nonumber \\
f_2(r_h)\left|L=2,J=3/2,F=3/2,F_z\right\rangle,\nonumber
\end{align}
where $L$ is the orbital angular momentum of the hole
  envelope function, $J$ is the hole spin, and the functions $\left|L,
    J, F, F_z\right\rangle$ are 
  eigenfunctions of the total angular momentum $\mathbf F = \mathbf L
  +\mathbf J$. Radial functions $f_l(r)$ ($l=0$, $2$) can be expressed
  in the spherical approximation 
  as~\cite{goupalov98} 
\begin{align}
f_l(r) &= \nonumber \\
&C \left[j_l(\zeta r/R) + (-1)^{l/2}
  \frac{j_2(\zeta)}{j_2(\sqrt{\beta}\zeta)}j_l(\sqrt{\beta} \zeta r/R)
\right], \nonumber 
\end{align}
where $j_l(x)$ are the spherical Bessel functions, $\beta$ is the
light to heavy hole mass ratio in the bulk material, $\zeta$ is the
first root of $j_0(x)j_2(\sqrt{\beta}\zeta) + j_2(x)
j_0(\sqrt{\beta}\zeta)=0$ and $C$ is the normalization constant.



The geometry of the system under study is illustrated in
Fig.~\ref{scheme}(a). In what follows we choose light propagation axis
to be $z \parallel [001]$. The optical selection rules at the trion
resonant excitation are similar to those for the interband absorption
in bulk material~\cite{OptOr}: under $\sigma^+$ polarized pulse action
the trions with $F_z=3/2$ and $1/2$ are formed, while for $\sigma^-$
pulse the trions with $F_z=-3/2$ and $-1/2$ are
formed~\cite{efros92}. Namely, under $\sigma^+$
light pulse optical transitions from the electron state with spin
projection $S_z=+1/2$ state to the $F_z=+3/2$ trion state and from
$S_z=-1/2$ electron state to the $F_z=+1/2$ trion state take place as
schematically shown in Fig.~\ref{scheme}(b). Similar selection rules
with inversion of $F_z$ and $S_z$ signs are relevant for the
$\sigma^-$ pump pulse. Hence, for given pump pulse helicity two
optical transitions are involved in contrast with the quantum disk
case with simple valence band, considered
in~\cite{yugova09,glazov:review}. 

To describe qualitatively the resident electron spin polarization
induced by the pump pulse we note that the probabilities to form a
trion from $S_z=+1/2$ and $S_z=-1/2$ initial states are
different. Indeed, as follows from the symmetry of the system, the
ratio of the matrix elements absolute values describing transitions to
the states with $|F_z|=3/2$ and $|F_z|=1/2$ equals to
$\sqrt{3}$~\cite{OptOr}. Therefore, for example, for $\sigma^+$ pump
the trions are formed more efficiently from $S_z=+1/2$ electrons than
from $S_z=-1/2$ ones. Under assumption that hole-in-trion spin
relaxation time, $\tau_s^T$, is much smaller than the trion lifetime,
$\tau_{QD}$, the electron returning after the trion recombination is
unpolarized. As a result the imbalance of $S_z=+1/2$ and $-1/2$
electrons occurs. This model explains the principle of spin
orientation of resident electrons in NCs and it is valid only for
relatively weak pump pulses. The microscopic theory for arbitrary pump
pulse powers is put forward below.  

In what follows we assume that the duration of the pump pulse,
$\tau_p$, is
small enough to neglect the spin precession in an external magnetic
field and the recombination and relaxation processes during the pump
pulse action: $\tau_p \ll \tau_{QD},\tau_s^T,T_L,\tau_s$, where
$\tau_s$ is the electron spin relaxation time in the quantum dot and
$T_L$ is the Larmor spin precession period. Hence, it is
enough to determine the transformation of electron spin by the pump
pulse, and solve afterwards kinetic equations for electron spin dynamics in the
interval between the pump pulses~\cite{yugova09}. Accordingly, we introduce
the six-component wavefunction $\hat\Psi$ of the quantum dot:
\begin{equation}
\hat{\Psi} =
[\psi_{+1/2},\psi_{-1/2},\chi_{+3/2},\chi_{+1/2},\chi_{-1/2},\chi_{-3/2}],
\end{equation}
where $\psi_{\pm1/2}$ refer to the resident electron spin states, and
four components $\chi_{F_z}$ ($F_z = \pm 3/2$, $\pm 1/2$)
refer to the corresponding trion states. It is assumed that the trion
states are chosen in the canonical basis for $\Gamma_8$ representation
of $T_d$ point symmetry group. As described above, the pump 
pulse with given helicity induces transitions between two electron and
two trion states [Fig~\ref{scheme}(b)]. Consequently only
four components  of $\hat\Psi$, namely, two electron ones,
$\psi_{\pm 1/2}$ and two corresponding trion once will be
modified by the pump pulse. The
non-stationary Schr\"{o}dinger equation can be
  written for the $\sigma^+$ polarized pump pulse as
(cf. Refs.~\cite{yugova09,glazov:review})
\begin{subequations}
\begin{equation}
\i\dot{\chi}_{+3/2} = \omega_0 \chi_{+3/2} + \sqrt{3} f(t) \e^{-\i \omega_p t} \psi_{+1/2},
\end{equation}
\begin{equation}
\i\dot{\psi}_{+1/2} = \sqrt{3} f(t) \e^{\i \omega_p t} \chi_{+3/2},
\end{equation}
\begin{equation}
\i\dot{\chi}_{+1/2} = \omega_0 \chi_{+1/2} + f(t) \e^{-\i \omega_p t} \psi_{-1/2},
\end{equation}
\begin{equation}
\i\dot{\psi}_{-1/2} = f(t) \e^{-\i \omega_p t} \chi_{+1/2}.
\end{equation}
\label{Schr}\end{subequations}
Here $f(t)$ is a smooth envelope of pump pulse defined as
\begin{equation}
f(t)=-\frac{\e^{i\omega_{p}t}}{\hbar} D E_{\sigma_+}(t),
\label{ft}
\end{equation}
where  $E_{\sigma_+}(t)\propto \e^{-\i\omega_pt}$ is the
right circularly polarized component of the electric field of the
pump and 
\begin{equation}
\label{dipole}
D=-\i\frac{ep_{cv}}{\omega_0m_0}\frac{1}{\sqrt{4\pi}}\int f_e(r)f_0(r)
\d\mathbf r
\end{equation}
is the  dipole matrix element~\cite{efros92} of interband
transition between the states $\psi_{-1/2}$ and $\chi_{+1/2}$. In the
last expression $e=-|e|$ is the electron charge, $m_0$ is the free
electron mass, and $p_{cv}$ is the interband matrix element of the
momentum operator taken between the conduction and valence band Bloch
functions at the $\Gamma$ point of the Brillouin zone. 
The dependence of electromagnetic field on coordinates is neglected in
Eq.~\eqref{ft} since the NC radius is much smaller than the radiation
wavelength. Similar to Eqs.~\ref{Schr} set of equations holds for
$\sigma^{-}$ pump pulse, in which case spins of hole $F_z$ and
electron $S_z$ should be inverted.

In  pump-probe experiments the generation and detection of electron
spin coherence is usually carried out by the train of pump (and probe)
pulses following with the repetition period $T_R$. It exceeds by far
the trion lifetime, $T_R \gg \tau_{QD}$. Therefore by the next pump
pulse the trion state in the NC is empty, $\chi_{F_z}(t\to
-\infty)\equiv 0$. However, the electron can be, in general, spin
polarized. Following Ref.~\cite{yugova09} we represent the solution of
the system~\eqref{Schr} after the pulse is over ($t \gg \tau_p$), as 
\begin{subequations} 
\label{Qpm5}
\begin{equation} 
\psi_{+1/2}(t\to+\infty) = Q_+\e^{\i \Phi_+}\psi_{+1/2}(t\to-\infty),
\end{equation} 
\begin{equation}
\psi_{-1/2}(t\to+\infty) = Q_-\e^{\i \Phi_-}\psi_{-1/2}(t\to-\infty).
\end{equation}
\end{subequations} 
Here $Q_{\pm}\in [0;1]$ and $\Phi_{\pm}\in (-\pi;\pi]$ are the
parameters, which depend on the shape, power and carrier frequency of
the pump pulse. The key difference of the present result,
Eqs.~\eqref{Qpm5}, from the case of simple valence band, considered in
Ref.~\cite{yugova09,glazov:review} is the fact that the both electron
spin components are transformed under the pump pulse
action. Equations~\eqref{Qpm5} allow one to relate the electron spin
components before the pulse, $\mathbf{S}^-=(S_x^-,S_y^-,S_z^-)$, and
after the pulse, $\mathbf{S}^+=(S_x^+,S_y^+,S_z^+)$, as follows: 
\begin{subequations}
\label{Spm}
\begin{equation} 
S_z^+ = \frac{Q_+^2-Q_-^2}{4} +\frac{Q_+^2+Q_-^2}{2}S_z^-,
\label{Sz1}
\end{equation} 
\begin{equation} 
S_x^+ = Q\cos{\Phi} S_x^- +Q\sin{\Phi} S_y^-, 
\label{Sx1}
\end{equation} 
\begin{equation} 
S_y^+ = Q\cos{\Phi} S_y^- - Q\sin{\Phi} S_x^-, 
\label{Sy1}
\end{equation}
\end{subequations} 
where $Q=Q_+Q_-$ and $\Phi=\Phi_+-\Phi_-$. It follows from Eqs.~\eqref{Spm}, that the electron spin pseudovectors before the pump pulse and after it are connected linearly. In accordance with Eq.~\eqref{Sz1} the pump pulse generates $z$-spin component (first term) and transforms the already present one (second term). The spin components in the plane perpendicular to the light propagation axis are reduced due to the factor $Q$ in Eqs.~\eqref{Sx1}~and~\eqref{Sy1} and are rotated around $z$-axis by the angle $\Phi$. Similar to Eq.~\eqref{Spm} relations hold also for $\sigma^-$ pump pulse, in which case $Q_+(\Phi_+)$ and $Q_-(\Phi_-)$ should be swapped.

To simplify the following discussion, we assume that the
  hole-in-trion spin relaxation is fast as compared with the radiative
  lifetime of the trion. In this case, the carrier returning after the
trion recombination is completely depolarized and the long-living spin
coherence generation is governed by Eqs.~(\ref{Spm}). Otherwise, to
determine the resident electron spin induced by the pump pulse one has
to solve the full system of spin dynamics equations taking into
account both electron and trion spins, cf. Refs.~\cite{zhu07,sokolova09,yugova12}.

\begin{figure}[t]
\includegraphics[width=\linewidth]{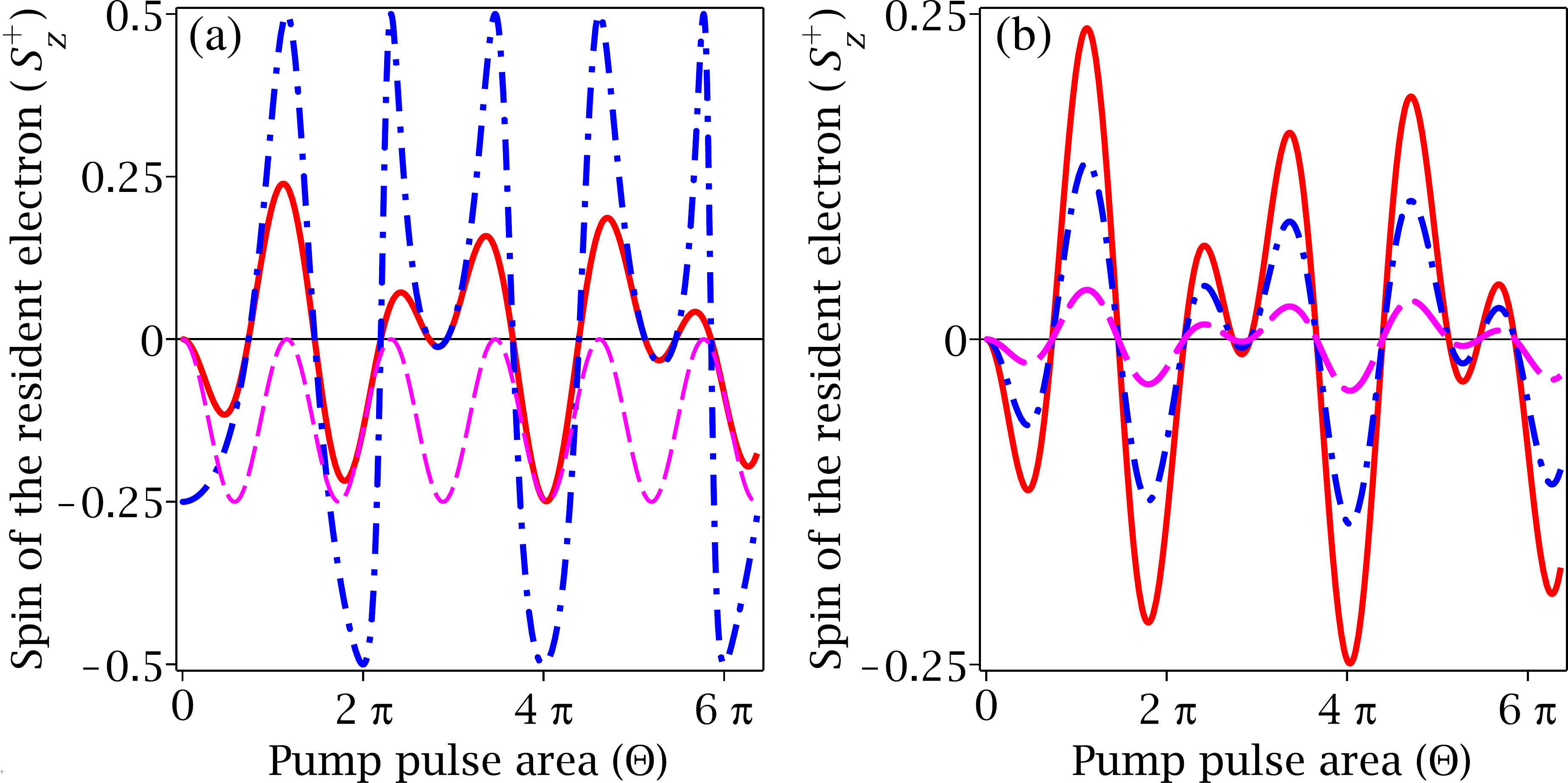}
\caption{
(a) The spin of the resident electron in the NC generated after a
single $\sigma^+$ pump pulse (red/solid line) and after the train of such
pulses (blue/dot-dashed line) as functions of the pump pulse area $\Theta$
at zero magnetic field. Resident electron spin in the quantum disk
after a single pulse $S_z=-\sin^2(\sqrt{3}\Theta/2)/4$ is presented by
thin magenta/dashed line. 
(b) Resident electron spin after a single Rosen and Zener pump pulse
for different detunings between the pump optical frequency and NC
transition frequency [$(\omega_0-\omega_p)\tau_p/(2\pi)=0,0.25,0.5$
for red/solid, blue/dash-dotted and magenta/dashed curves,
respectively]. 
}\label{SzFig}
\end{figure}

Let us consider an important limiting case of resonant pump pulse, where $\omega_p=\omega_0$. It can be shown, that $\Phi_\pm=0$ and $Q_\pm$ are expressed via the effective pump area $\Theta = 2 \int_{-\infty}^{\infty} f(t) \d t$ \cite{glazov:review} as
\begin{equation} 
Q_+ = \cos \left( \frac{\sqrt{3}\Theta}{2} \right), \quad Q_- = \cos \left( \frac{\Theta}{2} \right),
\end{equation} 
The periodic dependence of $Q_+$ and $Q_-$ on $\Theta$ is related to
the Rabi oscillations taking place in two-level systems corresponding
to optical transitions between electron state with $S_z=+1/2$ and the
trion state with $F_z=+3/2$, and between the electron state $S_z=-1/2$
and $F_z=+1/2$ trion.  The periods of these oscillations are
incommensurable due to the irrational ratio of the corresponding
matrix elements ($\sqrt{3}$). It results in the non-periodic
dependence of the electron spin $z$ component generated by the single
pump pulse
$S_z^{(1)}=\left[\sin^2(\Theta/2)-\sin^2(\sqrt{3}\Theta/2)\right]/4$
on pulse area (cf. Ref.~\cite{Rabi-Binder}). This dependence is shown
by the solid line in Fig.~\ref{SzFig}(a). For comparison, the dashed
line in Fig.~\ref{SzFig}(a) shows the electron spin $z$ component in
the model of simple valence band structure,
$-\sin^2(\sqrt{3}\Theta/2)/4$~\cite{yugova09}. In both cases $|S_z|
\le 1/4$, because in the ``best'' case, one of two-level systems gets
into the excited state, while the other one remains in the ground
state~\cite{yugova09}. Interestingly, that for the spherical
  NC considered here, depending on the pump pulse area $\Theta$ either
the transition related with electron with $S_z=+1/2$ or the one with
$S_z=-1/2$ can dominate. Hence, the electron spin after the pump
pulse with given helicity can be directed parallel or antiparallel to
$z$ axis depending on $\Theta$ by contrast to the self-organized
quantum dots with the simple valence band. 

Now let us briefly consider the case of non-resonant pumping. In order to obtain an analytic solution we consider the Rosen and Zener shape of the pump pulse:
\[ 
f(t) = \frac{\mu}{\cosh(\pi t / \tau_p)},
\] 
where the parameter $\mu$ determines the pulse area $\Theta=2\mu \tau_p$. Following Ref.~\cite{yugova09} we obtain for $Q_\pm$ and $\Phi_\pm$:
\begin{equation}
Q_\pm=\sqrt{1-\frac{\sin^2(\Theta_\pm/2)}{\cosh^2(\pi y)}},
\label{Qpm}
\end{equation}
\begin{equation}
\Phi_\pm=\arg\left\lbrace\frac{\Gamma^2\left(\frac{1}{2}-\mathrm iy\right)}
{\Gamma\left(\frac{1}{2}-\frac{\Theta_\pm}{2\pi}-\mathrm iy\right)
\Gamma\left(\frac{1}{2}+\frac{\Theta_\pm}{2\pi}-\mathrm iy\right)}\right\rbrace,
\end{equation}
where $\Theta_+ = \sqrt{3}\Theta_- \equiv \sqrt{3}\Theta$, and $y=(\omega_p-\omega_0)\tau_p/(2\pi)$ is the dimensionless detuning. It follows from Eqs.~\eqref{Sz1} and~\eqref{Qpm} that the electron spin $z$ component induced by a single detuned pump pulse has the form
\begin{equation}
S_z^{(1)}(\Theta,y)=\frac{S_z^{(1)}(\Theta,0)}{\cosh^2\left(\pi y\right)}=\frac{\sin^2(\Theta/2)-\sin^2(\sqrt{3}\Theta/2)}{4\cosh^2\left(\pi y\right)},
\label{S1} \end{equation}
and decreases monotonously with an increase of the detuning, $|y|$. This dependence is illustrated in Fig~\ref{SzFig}(b), where different curves correspond to different values of $y=0,0.25,0.5$. Overall dependence of $S_z^{(1)} (\Theta,y)$ is the same for all detunings in the case of Rosen and Zener pump pulse.

It is worth noting that for the deformed NC of elliptic shape or in NC
made of the wurzite semiconductor, the transitions involving
$S_z=+1/2$ and $S_z=-1/2$ electrons are characterized by different
detunings, in general. It results in even more complex dependence of
$S_z^{(1)}$ on the pump pulse area. In the particular case, where the splitting
between $|F_z|=3/2$ and $|F_z|=1/2$ trion states exceeds by far the
spectral width of pump pulse $\hbar/\tau_p$ only one optical
transition can be excited and spin orientation is described by the
model of~\cite{yugova09}. In the opposite case the model presented here is valid. Hence, by choosing appropriate
  pulse spectral width and strength one can orient spin in
  arbitrary direction for nanocrystal systems like studied in Refs.~\cite{doi:10.1021/nl801057q,PhysRevB.84.085304}.

\section{Spin accumulation caused by the train of pump pulses}
\label{sec3}

In the pump-probe Faraday and Kerr rotation experiments the sample is
usually subjected to a train of pump pulses that 
follow with a certain repetition period $T_R$. We assume
  that $T_R$ exceeds by far the radiative lifetime of a trion in a
  quantum dot, $\tau_{QD}$, but it may be comparable or smaller than
than the single electron spin relaxation time in a QD, $T_R \le
\tau_{s,e}$. In this case, the electron spin  retains the
  memory of being exposed to the previous pulses, resulting in the
  resonant spin amplification and spin mode-locking
  effects~\cite{Kikkawa98,A.Greilich07212006}.

Let us consider firstly the simple case, where the external
  magnetic field is absent,  $\mathbf B=0$. For simplicity we assume also
  that the electron spin relaxation time exceeds by far the pump pulse
repetition period. Hence, electron spin is conserved between the pump
pulses and $\mathbf S^+ = \mathbf S^-$ in the steady state. As a result for the electron spin components after
the sufficiently long train of the pump pulses we obtain
\begin{subequations}
\label{Sinf:0}
\begin{equation}
S_x^{+} = S_x^{-} = S_y^{+} = S_y^{-} = 0,
 \end{equation}
\begin{equation}
S_z^{+} = S_z^{-} = \frac{Q_+^2-Q_-^2}{4-2(Q_+^2+Q_-^2)}.
 \end{equation}
\end{subequations}
The  dependence of electron spin $S_z$ component accumulated
  by the train of the pump pulses and calculated after
  Eqs.~(\ref{Sinf:0}) is shown in Fig.~\ref{SzFig}(a) by the
  dash-dotted line. Interestingly, that for the Rosen and Zener pump
  pulse shape, the dependence of $S_z$ on the pump pulse area is
  independent of the detuning between the pump pulse frequency and the
  quantum dot transition frequency, as follows from Eqs.~(\ref{Qpm})
  and (\ref{Sinf:0}). We have checked that this dependence is weak for rectangular-shaped pulse.

It is noteworthy that the electron spin
  polarization may reach 100\% for spherical NC. This is in contrast
  to the classical 
  optical orientation regime in the systems with $\Gamma_8$ symmetry
  of the valence band, where up to $50\%$ spin polarization can be
  realized. Such an enhancement of spin polarization and the
  dependence of the electron spin direction on the pump pulse area is
  related with the two-level nature of the optical transitions in NCs.
For example, if one of the two-level systems (associated with the
$F_z=+3/2$ or $F_z=+1/2$) trion is inactive, which happens if
$Q_+=1$ or $Q_-=1$, only electrons with the fixed spin component are depolarized
    ($S_z=1/2$ or $-1/2$, respectively). Such a situation is similar
    to the one considered in Refs.~\cite{greilich06,yugova09}. As a
    result, the sufficiently long 
    train of pump pulse brings  completely erases one electron spin
    component, hence, resident carriers become fully
    polarized.

In the general case, where the magnetic field is applied to
  the NC, the resident electron spin
  dynamics is governed by the following equation:
\begin{equation}
\frac{\mathrm d \mathbf S}{\mathrm dt} +\mathbf{S}\times
\mathbf \Omega_L+\frac{\textbf{S}}{\tau_{s,e}} = 0.
\label{precession} \end{equation}
The equation takes into account the electron spin relaxation
processes and the precession of electron
spins in the magnetic field $\textbf{B}$ with the frequency
  $\mathbf \Omega_L = 
g_e\mu_B \mathbf B$, where $g_e$ is the electron Land\'{e} factor and $\mu_B$
is the Bohr magneton. Following
  Refs.~\cite{yugova09,yugova12} we obtain the following steady-state
  expressions  for the electron spin
  components $\mathbf S^-$ right before the pump pulse arrival
\begin{subequations}
\label{Sxyzp}
\begin{equation}
S_x^- = K S_y^-,
\end{equation}
\begin{equation}
S_y^- = \frac{Q_-^2-Q_+^2}{4\Delta}\e^{-T_R/\tau_{s,e}}\sin(\Omega_L T_R),
\end{equation}
\begin{align}
S_z^- &= \frac{Q_-^2-Q_+^2}{4\Delta}\e^{-T_R/\tau_{s,e}} \times \\
&\left[ Q(\cos\Phi-K\sin\Phi)\e^{-T_R/\tau_{s,e}}-\cos(\Omega_L
  T_R)\right], \nonumber
\end{align}
\end{subequations}
where
\begin{widetext}
\begin{subequations}
\begin{equation}
\Delta = 1 - \e^{-T_R/\tau_{s,e}}  \left[ \frac{Q_+^2+Q_-^2}{2} + Q(\cos\Phi-K\sin\Phi) \right]\cos(\Omega_L T_R)
+\frac{Q(Q_+^2+Q_-^2)}{2}\e^{-2T_R/\tau_{s,e}}(\cos\Phi-K\sin\Phi),
\label{eqDelta} \end{equation}
\begin{equation}
K = \frac{Q\e^{-T_R/\tau_{s,e}}\sin\Phi}{1-Q\e^{-T_R/\tau_{s,e}}\cos\Phi}.
\label{eqK}
\end{equation}
\end{subequations}
\end{widetext}
We remind that $Q=Q_+Q_-$. In the limit of 
$Q_-=1$, i.e. where one of the two level systems is inactive Eqs.~(\ref{Sxyzp}) pass
  to those obtained in Ref.~\cite{yugova09} for the simple valence band system. 


In order to analyze Eqs.~\eqref{Sxyzp} we consider an important limiting case of
resonant pumping, where $\Phi=0$, and neglect
completely the resident electron spin relaxation
($\tau_{s,e}\to\infty$). This results in $K=0$,
\begin{equation}
\Delta = 1-\left[\frac{Q_+^2+Q_-^2}{2} + Q \right]\cos \Omega_L T_R + \frac{Q(Q_+^2+Q_-^2)}{2},
\end{equation}
see Eqs.~(\ref{eqDelta}), (\ref{eqK}), and leads to the
  following simplified expressions for the electron spin components
  accumulated by the train of the pump pulses:
\begin{subequations}
\label{S:train:simple}
\begin{equation}
 S_x^-=0,
\end{equation}
\begin{equation}
S_y^-=\frac{Q_-^2-Q_+^2}{4\Delta}\sin \Omega_L T_R,
\end{equation}
\begin{equation}
S_z^-=\frac{Q_-^2-Q_+^2}{4\Delta}\left[Q-\cos \Omega_L T_R \right].
\end{equation}
\end{subequations}
Clearly, $S_x \equiv 0$ for the considered geometry, since
  the magnetic field is parallel to $x$ axis and optical pulses do not
  lead to the spin rotation in $(xy)$ plane in the resonant case.

Under the phase synchronization
condition~\cite{A.Greilich07212006,A.Greilich09282007,Kikkawa98,carter:167403,yugova11,yugova12,glazov08a},
\begin{equation}
\label{PSC} 
\Omega_L T_R = 2\pi N, \quad N=0, 1, 2, \ldots ,
\end{equation}
  the spin of electron makes an integer number of turns between the
  subsequent pump pulses and the electron spin is given by the same
  expressions as in the absence of the magnetic field, see
  Eqs.~(\ref{Sinf:0}) and may reach $\pm 1/2$ depending on pump pulse area.

For weak pump pulses, where the pulse area $\Theta \ll 1$ one obtains
\begin{align}
&S_z^- \approx \frac{\Theta^2}{16} \times \nonumber\\
&\left(1-\frac{(2{+}\cos{\Omega_LT_R})\Theta^2}{3
  [(1-\cos{\Omega_L
  T_R})(2-\Theta^2) + (41-17\cos{\Omega_LT_R})\frac{\Theta^4}{96}]}\right), 
\label{Theta0}
\end{align}
\[
{S_y^- \approx  \frac{\Theta^2}{8} \frac{ \sin{\Omega_L
    T_R} }{(1-\cos{\Omega_L
  T_R})(2-\Theta^2) + (41-17\cos{\Omega_LT_R})\frac{\Theta^4}{96} } }.
\]
Note, that the peaks of the dependence of $S_z$ on magnetic
  field corresponding to the phase synchronization
  condition~\eqref{PSC} are very sharp. If the spin
    relaxation and the detuning are completely neglected, their width
    is determined by 
    the pump pulse area ($\sim\Theta^2$), otherwise it is
    determined by ratio $T_R/\tau_{s,e} \ll 1$ or $|\Phi|$,
    whichever is larger, see
  Ref.~\cite{yugova09} for details. In the resonant case the electron
  spin $z$-component equals to $-1/4$ like in the bulk system with
  $\Gamma_8$ symmetry since a train of weak pulses is similar to the
  \emph{cw} pumping of the bulk material.

In the other important case, where $Q_+$ or $Q_-=0$ the corresponding
two-level system 
  is excited to the trion state with a unit probability. In this
limit $Q=0$, and the electron spin components take very
  simple form:
\begin{subequations}
\begin{equation}
S_x^- = 0, \quad S_y^- = -\frac{Q_+^2}{4\Delta}\sin \Omega_L T_R,
\end{equation}
\begin{equation}
S_z^- = \frac{Q_+^2}{4\Delta}\cos \Omega_L T_R,
\end{equation}
\end{subequations}
where $\Delta = 1-\cos{(\Omega_LT_R)} Q_+^2/2$ and we assumed
  that $Q_-=0$ for the sake of definiteness. In this situation
  under the synchronization condition~(\ref{PSC}) the electron spin
  becomes fully polarized.


\section{Probing the electron spin in quantum dots}
\label{sec4}

\subsection{Formation of spin Faraday and ellipticity signals in NC}

The detection of the QD spin polarization in pump-probe Faraday and
Kerr rotation experiments is carried out by a weak linearly polarized
probe pulse. The electric field of the probe pulse oscillates along
the $x$ axis and can be written as 
\begin{equation}
\textbf{E}^{pr}(\textbf{r},t)=E_x^{pr}(\textbf{r},t)\mathbf{o}_x +c.c.
\nonumber \end{equation}
Here we assume that
$E_x^{pr}(\textbf{r},t) = E_0 s(t) \e^{-\i\omega_{pr}t}$, where
$\omega_{pr}$ is the carrier frequency of the probe
  beam, $E_0$ is the probe pulse amplitude, $s(t)$ is its smooth envelope, and
$\textbf{o}_i$ ($i=x,y,z$) is the unit vector along $i$th
  axis. The Faraday signal
  $\mathcal F$ detected in the transmission geometry can be written as~\cite{yugova09,glazov:review}
\begin{equation}
\label{spinF}
\mathcal F = \lim_{z\to + \infty} \int
\left[\left|E^{(t)}_{x'}(z,t)\right|^2 -
  \left|E^{(t)}_{y'}(z,t)\right|^2 \right]
\mathrm dt,
\end{equation}
where $x'$, $y'$ axes are oriented at  $45^{\circ}$ with respect to the
initial reference frame $x$, $y$; $E^{(t)}_{x'}(z,t)$ and $E^{(t)}_{y'}(z,t)$
are the components of the transmitted field. Kerr rotation signal can
be presented similarly to Eq.~(\ref{spinF}) with the replacement of
the transmitted fields by the reflected ones. Similarly, the
ellipticity of the transmitted beam reads
\begin{equation}
\label{spinE}
\mathcal E = \lim_{z\to + \infty} \int
\left[\left|E^{(t)}_{\sigma^+}(z,t)\right|^2 -
  \left|E^{(t)}_{\sigma^-}(z,t)\right|^2 \right]
\mathrm dt.
\end{equation}
Here the subscripts $\sigma^+$ and $\sigma^-$ correspond to the
circular components of the transmitted light. 

\begin{figure*}[hptb]
\includegraphics[width=0.9\textwidth]{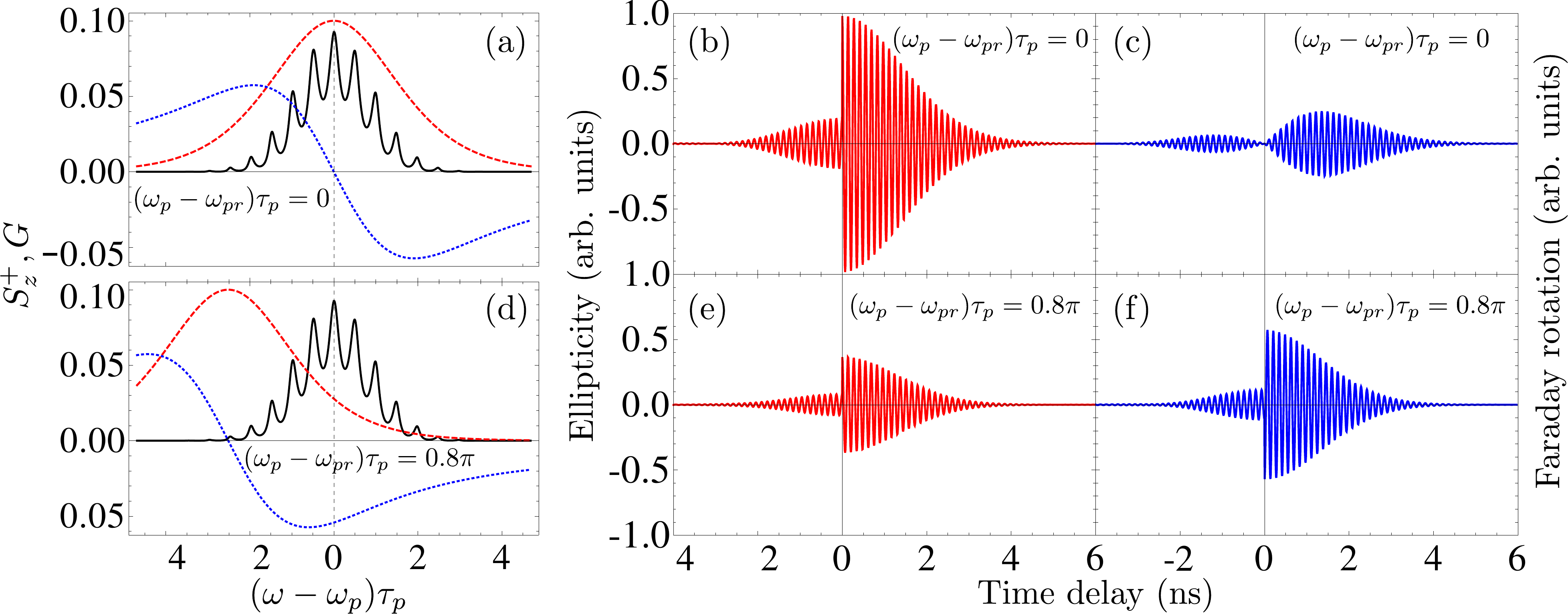}
\caption{
(a) Electron spin polarization, $S_z^+$, created by a train of
Rozen \& Zener $\sigma^+$ polarized pump pulses with the repetition period
$T_R=13.2$~ns, as a function of NC optical transition frequency $\omega_0$
(black/solid line). The real (red/dashed) and imaginary 
(blue/dotted) parts of function $G(\omega-\omega_p)$ are shown in
panels (a) and (d). Panels (b,c) represent the time resolved dependence of the
spin ellipticity and Faraday rotation signals in the QD ensemble for
the degenerate pump-probe regime ($\omega_p = 
\omega_{pr}$). Panels (d) -- (f): same as in (a) -- (c), but for the
nondegenerate regime 
$(\omega_p-\omega_{pr})\tau_p=0.8\pi$.
Calculations are carried out for the pump pulses of the
area $\Theta=\pi$, the in-plane magnetic field $B=1$~T, $\tau_p=100$~fs,
$\tau_{s,e}=1$~$\mu$s, $\hbar\bar{\omega}_0 = 1.4$~eV, $\hbar\Delta\omega_0 =
  6.15$~meV~\cite{A.Greilich07212006}. Parameters $A$ and $C$ in
  Eq.~(\ref{g:eff}) are: $A =
  -1.75$~eV$^{-1}$, $C = 2.99$~\cite{A.Greilich07212006}.
}\label{ModeLock}
\end{figure*}

In order to find the transmitted and reflected
  electromagnetic field we follow the procedure developed in
  Refs.~\cite{yugova09,glazov:review} and calculate response of the NC
  to the linearly polarized probe field. To that end, the electric
  field of the probe pulse is decomposed in a superposition of
  circularly polarized waves and
  system~(\ref{Schr}) and its counterpart for other circular
  polarization are used to determine the change of the quantum dot
  wavefunction. Corresponding equations are solved in the first order
  in electric field $\mathbf E^{pr}$~\footnote{
Linearly polarized pulse affects the electron spin
  polarization in the nanocrystal~\cite{zhukov10}. One can show
that electron spin components before ($\mathbf S^-$) and after
($\mathbf S^+$) the linearly polarized pulse are related as
$\mathbf{S}^+ = Q_l \mathbf{S}^-$, where, e.g. for resonant pulse,
$Q_l =\cos^{2}{(\Theta_l/\sqrt{2})}$, there $\Theta_l$ is the linearly
polarized pulse area.
}. Afterwards the dielectric
  polarization of the NC is calculated and the re-emitted field is
  determined from the Maxwell equations. The resulting expressions for
spin Faraday and ellipticity signals for the three-dimensional (bulk)
array of NCs read:
\begin{align}
\label{signals}
&\mathcal{E}+\i\mathcal{F} = \\
&\frac{\pi N_{NC}
  L}{q^2\tau_{NC}}  |E_0|^2 \left[J_z+\i\left\langle\left\lbrace
    J_x,J_y\right\rbrace_s\right\rangle-2S_z\right]G(\omega_{pr}-\omega_0). 
\nonumber   
\end{align}
Here $q=\omega_{pr}\sqrt{\varepsilon_b}/c$ is the radiation wavevector
in the medium, $\varepsilon_b$ is
the dielectric constant of the matrix, which is assumed to coincide
with the background dielectric constant of the NCs, 
\begin{equation}
\tau_{NC} = \frac{3\hbar c^3}{4|D|^2\omega_0^3\varepsilon_b} \nonumber
\end{equation}
is the radiative lifetime of an electron-hole pair confined in the
NC \cite{efros92},
$N_{NC}$ is the (volume) density of the NC, $L$ is the thickness of the
{array}. Complex-valued function $G(\omega_{pr}-\omega_0)$ in Eq.~(\ref{signals})
determines the spectral sensitivity of the spin Faraday and
ellipticity signals~\cite{yugova09,glazov:review}
\[
G(\Lambda) =\int_{0}^{\infty}\d t\e^{\i\Lambda
  t}\int_{-\infty}^{\infty}\d t' s(t')s(t+t'). 
\]
The explicit expressions for $G(\Lambda)$ are given by
  Eq. (61) of Ref.~\cite{yugova09} for different probe pulse
  shapes. Its real and imaginary parts {for the case of Rozen \& Zener
    pulse} are shown by red {dashed} and blue {dotted} lines in
  Fig.~\ref{ModeLock}(a), respectively. It is noteworthy, that the
  spin Kerr signal is a superposition of spin Faraday and ellipticity
  signals with the coefficients determined by phase acquired by the
  probe pulse in the cap layer~\cite{glazov:review,yugova09}.

Quantities $S_z$ and $J_z$ in Eq.~(\ref{signals}) are the
  electron and the hole-in-trion $z$-spin components at the moment of
  the probe pulse arrival. The probe pulse duration
is assumed to be short as compared with all other time scales in the
  system,
and therefore, the electron spin can be considered as frozen during the probe pulse action. Interestingly, the Faraday rotation and
  ellipticity signals contain contribution from the quantity $\langle
  \left\lbrace J_x,J_y\right\rbrace_s\rangle$, which is the quantum mechanical
  average of the symmetrized product of $J_x$, $J_y$ hole spin
  operators, $\{  J_x,J_y\}_s = (J_xJ_y + J_yJ_x)/2$. Note, that the
  similar combination of electron spin operators (although being
  symmetry allowed) does not contribute to the spin Faraday and ellipticity
  effects since second powers of electron spin operators reduce to the
first powers~\footnote{In the linear in the $\mathbf E^{pr}$ regime the
  probe-induced dielectric polarization does not contain contributions
  like $\{ J_x,\sigma_y\}_s$ as well.}
The contributions due to
  the trion spin polarization vanish if the pump-probe delay exceeds
  $\tau_{NC}$, the radiative lifetime of an electron-hole pair is the
  nanocrystal, therefore pump-probe signals at long enough delays
  are determined by the electron spin $z$ component only.
It is assumed below that the pump-probe delay exceeds the
  trion lifetime in the NC and we focus solely on the electron spin
contribution to the Faraday and ellipticity signals. 

We note, that due to the degeneracy of the heavy- and light-hole states the spin Faraday, Kerr and ellipticity signals are
proportional to the electron spin projection onto the light
propagation axis. Hence, by changing the direction of light
propagation one can study the dynamics of all spin pseudovector
components, making spherical nanocrystals most suitable for the spin
state tomography measurements~\cite{Kosaka2009}.

Before we proceed with the discussion of spin Faraday and ellipticity
signals temporal behavior, let us estimate the typical spin signal
strengths. For the NC array with a concentration $N_{NC} =
10^{15}$~cm$^{-3}$, thickness $L=10^{-4}$~cm the Faraday rotation
angle can be estimated  
  according to Eq.~(\ref{spinF}) as $\sim 10$~mrad for $\tau_{NC} =
  400$~ps, $\hbar\bar{\omega}_0 =1.4$~eV,  $\varepsilon_b=11$.

\subsection{Temporal dependence of spin Faraday and ellipticity signals}

In real NC ensembles the optical transition energies and electron Larmor precession frequencies vary. To simulate the spin signals of such inhomogeneous array of NCs we consider a normal distribution of resonant transition frequencies in NCs, $\omega_0$, around a certain mean value $\bar{\omega}_0$ with a dispersion $\Delta\omega_0$~\cite{yugova09}
\begin{equation}
\label{rho}
\rho(\omega_0) = \frac{1}{\sqrt{2\pi(\Delta\omega_0)^2}}\exp\left[-\frac{(\omega_0-\bar{\omega}_0)^2}{2(\Delta\omega_0)^2} \right].
\end{equation}
The electron spin precession frequency is determined both by
  the electron $g$-factor and by the fluctuations of nuclei spins, the
latter are neglected hereinafter for simplicity. The main contribution
to the $g$-factor spread is related with its variation with the band
gap, see Refs.~\cite{PhysRevB.75.245302,ivchenko05a,glazov2010a} for
details. Under the conditions that $\Delta\omega_0 \ll \bar\omega_0$ the
dependence of the $g$-factor on the resonance frequency can be taken
in the linear form,  
\begin{equation}
\label{g:eff}
g_e(\omega_0)=A\hbar\omega_0+C,
\end{equation}
where $A$ and $C$ are some coefficients.
In accordance with Eq.~\eqref{signals} the electron spin Faraday and ellipticity signals as functions of the
  time-delay between the probe and the pump pulses $t$ in
  the quantum dot ensemble are given by~(cf. Refs.~\cite{glazov2010a,glazov:review})
\begin{align}
\label{time:resolved}
&\mathcal{E}+\i\mathcal{F} = \\
& -2 \frac{\pi N_{NC}
  L}{q^2\tau_{NC}} |E_0|^2 \int
S_z(\omega_0,\omega_p;t)G(\omega_{pr}-\omega_0)\rho(\omega_0)\d\omega_0. \nonumber
\end{align}
where $S_z(\omega_0,\omega_p;t)$ is the temporal dependence of the
  electron spin in the quantum dot with the resonance frequency
  $\omega_0$ induced by the pump with the carrier frequency $\omega_p$, which
  can be found from Eqs.~\eqref{precession} and \eqref{Sxyzp}. The
  variation of the $\tau_{NC}$ with the variation of resonance
  frequency is neglected 
  in Eq.~\eqref{time:resolved} since this dependence is very
    smooth on the scale of the inverse pulse duration. 

The spread of spin precession frequencies caused by the
  $g$-factor spread gives rise to important consequences. First, for
  some quantum dots in the ensemble the spin precession frequency and
  the pump pulse repetition frequency are commensurable, see
  Eq.~\eqref{PSC}, resulting in 
  the strong enhancement of the $S_z$ in these dots. The distribution
  of the electron spin $z$-component right after the pump pulse
  arrival as a function of the quantum dot optical transition frequency
  is shown by the solid curve in Fig.~\ref{ModeLock}(a),(d).
  The peaks on these curve correspond to the dots which satisfy the
  phase synchronization 
  condition~\eqref{PSC}. This yields the
  mode-locking of electron spin precession~\cite{A.Greilich07212006}:
  the total spin of the ensemble induced by the pump pulse gets dephased due to the spread of the
  spin precession frequencies on the nanosecond time scale, and before the next pulse arrival the spin polarization of
  ensemble emerges due to the contribution of the nanocrystals where
  the spin precession is synchronized. This is clearly seen in
  Figs.~\ref{ModeLock}(b),(c),(e) and (f), where the temporal
  dependence of the spin ellipticity [(b), (e)] and Faraday rotation
  [(c), (f)] are presented for different detunings between the pump and
  probe pulses. It is worth to stress, that the allowance for the
  nuclei-induced electron frequency effect can result in even higher
  amplitudes of the signals at negative
  delays~\cite{A.Greilich09282007,carter:167403,yugova11}.

Second consequence of the $g$-factor spread is clearly
  manifested in Fig.~\ref{ModeLock}(c) where the Faraday rotation
  signal for degenerate pump and probe pulses is
  demonstrated. Notably, the Faraday rotation is absent at $t=0$, and
  the signal amplitude is a non-monotonous function of the time delay:
  firstly it grows with time and afterwards it decays. This is a feature
of the spectral sensitivity of the Faraday effect which is described
by the odd function of the detuning, see blue dotted curve in
Fig.~\ref{ModeLock}(a). The dependence of electron spin $z$-component
on {the} detuning, $\omega - \omega_p$, is symmetric right after the pump
pulse arrival ($t=0$) 
but becomes asymmetric as time goes by due to the correlation of
electron $g$-factor and the resonance frequency of the
NC. It results in the growth of the Faraday rotation signal with
time. For even higher time delays the Faraday rotation amplitude
decays due to the spin dephasing~\cite{glazov2010a}. The pump-probe
detuning {itself} introduces the asymmetry of the spin distribution
with respect to the probe frequency, see Fig.~\ref{ModeLock}(d) and
the Faraday rotation signal shows the behaviour similar to that of the
ellipticity, namely, damped oscillations. The interaction of electron
spins with nuclei may result in the breaking of the direct link of the
electron spin precession frequency and optical transition frequency
making the growth of spin Faraday signal less
pronounced.

\section{Conclusions}
\label{sec5}

In the present work the microscopic description of the
  resident electron spin dynamics under the pump-probe conditions in spherical
  nanocrystals is developed. It is shown, that the complex structure
of the valence band allows one to control the
magnitude and direction of the resident electron spin in the
nanocrystal by means of the pump power variation. It results from
  the possibility to activate or  suppress two possible optical
  transition paths, involving heavy and light holes. We have
  demonstrated that the periodic pumping can result in the pronounced
  spin coherence mode-locking and lead to the complete polarization of
  electron spins by sufficiently long trains of the pump pulses.

Spin Faraday and ellipticity signals were calculated for the
 three-dimensional inhomogeneous  arrays of nanocrystals. It was shown
 that the signals possess information on electron and hole spin
 dynamics, moreover, due to the ensemble inhomogeneity the temporal
 behavior of the Faraday and ellipticity effects can be different.

 The results of this work can be also applied to the description of
 spin coherence generation and detection for electrons in colloidal
 nanocrystals or for electrons localized on donors in the bulk
 GaAs-type semiconductors where the donor-bound exciton can be
 optically excited. 

\acknowledgements

We thank A. Greilich, E.L. Ivchenko, D.R. Yakovlev and I.A. Yugova  for valuable
discussions. Financial support of RFBR, RF President Grant 
NSh-5442.2012.2, and EU projects Spinoptronics and POLAPHEN is
gratefully acknowledged.

%

\end{document}